\documentclass[lettersize,journal]{IEEEtran}
\IEEEoverridecommandlockouts

\usepackage{cite}
\usepackage{amsmath,amssymb,amsfonts}
\usepackage{algorithmic}
\usepackage{graphicx}
\usepackage{textcomp}
\usepackage[dvipsnames, table]{xcolor}
\usepackage{paralist}
\usepackage[inline]{enumitem}
\usepackage{caption}
\usepackage{makecell}
\usepackage{tabu, booktabs}
\usepackage{float}
\usepackage{subcaption}
\usepackage{comment}
\usepackage{todonotes}
\usepackage[breaklinks,colorlinks,bookmarks=true]{hyperref}
\usepackage{csquotes}
\usepackage{soul}
\usepackage{pifont}
\usepackage{multirow}
\usepackage{tcolorbox}
\usepackage{tabularx}
\usepackage{array}
\usepackage{colortbl}
\usepackage{siunitx}
\usepackage{svg}

\definecolor{magma_darker}{HTML}{fdc38a}
\definecolor{magma_dark}{HTML}{e15666}
\definecolor{magma_light}{HTML}{82247f}
\definecolor{magma_lighter}{HTML}{1f0c43}

\definecolor[named]{xBlue}{HTML}{18647E}
\definecolor[named]{xOrange}{HTML}{FF9B00}
\definecolor[named]{xGray}{HTML}{808080}
\definecolor[named]{xGreen}{HTML}{60B950}
\definecolor[named]{xRed}{HTML}{A30B37}
\definecolor[named]{xDarkBlue}{cmyk}{1,0.58,0,0.21}

\definecolor{cadetblue}{rgb}{0.37, 0.62, 0.63} 
\definecolor{grayish}{rgb}{0.6, 0.6, 0.6} 
\definecolor{redorange}{rgb}{0.91, 0.41, 0.17} 
\definecolor{limegreen}{rgb}{0.5, 0.8, 0.2} 

\hypersetup{
    colorlinks=true,
    linkcolor=xOrange,
    filecolor=magenta,      
    urlcolor=xDarkBlue,
    citecolor=xBlue,
}

\usepackage[a4paper, total={184mm,239mm}]{geometry}
\def\BibTeX{{\rm B\kern-.05em{\sc i\kern-.025em b}\kern-.08em
    T\kern-.1667em\lower.7ex\hbox{E}\kern-.125emX}}
    
\ExplSyntaxOn
\DeclareExpandableDocumentCommand{\convertlen}{ O{cm} m }
 {
  \dim_to_decimal_in_unit:nn { #2 } { 1 #1 } cm
 }
\ExplSyntaxOff
\usepackage{gnuplottex}
\usepackage{tikz}
\usepackage{mathtools}

\usetikzlibrary{matrix,calc,positioning,arrows.meta}

\newcommand*\circled[1]{\tikz[baseline=(char.base)]{\node[shape=circle,fill,inner sep=0.5pt] (char) {\textcolor{white}{#1}};}}

\begin{document}

\title{CarbonCall: Sustainability-Aware Function Calling for Large Language Models on Edge Devices}

\author{
\IEEEauthorblockN{Varatheepan~Paramanayakam$^{1}$, Andreas Karatzas$^{1}$, Iraklis Anagnostopoulos$^1$, Dimitrios Stamoulis$^2$}~\\
\IEEEauthorblockA{$^1$School of Electrical, Computer and Biomedical Engineering, Southern Illinois University, Carbondale, IL, U.S.A.}
\IEEEauthorblockA{$^2$Department of Electrical and Computer Engineering, The University of Texas at Austin, Austin, TX, U.S.A.}
\IEEEauthorblockA{Email: \{varatheepan, andreas.karatzas, iraklis.anagno\}@siu.edu, dstamoulis@utexas.edu}
}

\maketitle
\begin{abstract}
Large Language Models (LLMs) enable real-time function calling in edge AI systems but introduce significant computational overhead, leading to high power consumption and carbon emissions. Existing methods optimize for performance while neglecting sustainability, making them inefficient for energy-constrained environments. We introduce CarbonCall, a sustainability-aware function-calling framework that integrates dynamic tool selection, carbon-aware execution, and quantized LLM adaptation. CarbonCall adjusts power thresholds based on real-time carbon intensity forecasts and switches between model variants to sustain high tokens-per-second throughput under power constraints. Experiments on an NVIDIA Jetson AGX Orin show that CarbonCall reduces carbon emissions by up to 52\%, power consumption by 30\%, and execution time by 30\%, while maintaining high efficiency.
\end{abstract}


\setlist{nosep}

\section{Introduction}

A key advancement in Generative AI is the integration of Large Language Models (LLMs) into agentic systems, enabling autonomous API interactions in response to user requests~\cite{paramanayakam2024less,patil2025gorilla}. This function-calling paradigm powers virtual assistants and edge AI, allowing LLMs to dynamically select and execute tools for structured data processing. However, function calling adds computational overhead, requiring LLMs to parse APIs, assess relevance, and generate structured requests, all within the tight resource and power limits of edge devices~\cite{yuan2024llm}.

Edge systems have limited power and computing resources, making LLM execution inefficient~\cite{yuan2024llm,karatzas2024mapformer}. Function calling increases energy demands further, as each query triggers additional inference cycles, leading to higher latency and power consumption. Unlike cloud servers, which optimize workloads using large-scale energy-efficient computing, edge devices rely on localized, often carbon-intensive power sources that vary by region~\cite{maji2022carboncast, wu2022sustainable}. As agentic LLMs become more common in real-time applications~\cite{patil2025gorilla}, their energy footprint raises sustainability concerns, requiring efficient execution without compromising responsiveness. Existing optimizations, such as replacing large models (e.g., 70B, 400B parameters) with smaller LLMs (e.g., 1.5B, 3.8B, 7B, 8B), improve efficiency but focus primarily on performance, often neglecting sustainability~\cite{yuan2024llm,paramanayakam2024less}. Addressing these challenges requires sustainability-aware function-calling strategies that reduce power consumption while maintaining responsiveness.

Reducing operational carbon emissions adds further complexity to an already difficult challenge. While lowering power consumption on edge devices is crucial for sustainability, simply minimizing energy use is not sufficient. Research has shown that focusing solely on power and energy efficiency does not always lead to a lower carbon footprint~\cite{gupta2022act}. Static power-saving techniques, such as reducing CPU and GPU frequencies, can decrease energy consumption but come at the cost of significant performance degradation, making them impractical for real-time function calling in LLMs. Existing approaches tend to prioritize either efficiency or responsiveness~\cite{yuan2024llm,paramanayakam2024less}, yet no current solutions effectively address both aspects simultaneously at the edge. A promising alternative, explored in other AI workloads, is the use of mixed-quality models, where models of the same architecture but with varying computational complexity dynamically adjust to resource constraints~\cite{li2023clover}. While this strategy has been applied to DNN inference~\cite{sankaranarayanan2024pulse}, it has not yet been adapted for LLM function calling at the edge, where balancing power efficiency and performance remains a critical challenge.

In this paper, we present \textbf{CarbonCall}, a sustainability-aware function-calling framework for LLMs on edge devices. Our approach addresses the challenges of energy efficiency, computational overhead, and operational carbon footprint in real-time function calling. Unlike existing methods that rely on offline training, auxiliary models (e.g., helper and recommender models), or static power-saving techniques, CarbonCall dynamically adapts tool selection, power usage, and model complexity to optimize performance while reducing energy consumption. \textbf{The core contributions of CarbonCall are threefold:}
\begin{inparaenum}
    \item[\circled{1}] A dynamic tool selection mechanism that eliminates the need for offline training or computationally intensive recommenders, enabling efficient function execution with minimal overhead.
    \item[\circled{2}] A carbon-aware execution strategy that adjusts device power consumption in real time based on carbon intensity (CI) variations, reducing operational emissions while maintaining system responsiveness.
    \item[\circled{3}] A mixed-quality LLM approach, where models with higher quantization levels are leveraged to ensure fast response times at lower hardware frequencies, significantly improving energy efficiency.
\end{inparaenum}    
By integrating these strategies, CarbonCall provides a sustainable and efficient solution for LLM-based function calling at the edge, addressing both performance and environmental aspects.
\section{Related Work}

Existing work tackles two important focus areas of this paper: \begin{inparaenum}[(\bgroup\bfseries i\egroup)]
\item  Edge LLM inference and function calling methods;
\item Carbon aware execution strategies.
\end{inparaenum} To enhance execution efficiency, EdgeLLM~\cite{xu2024edgellm} proposes a confidence mechanism to prune unlikely branches in the token tree and eliminate wasteful computations. Although speculative execution improves efficiency in general, its impact on function calling remains underexplored. 
To optimize function execution, caching-based approaches have been explored, such as query- and tool-caching \cite{singh2024llm}, which store frequently used LLM responses to reduce redundant computation.
STE~\cite{Wang2024LLMsIT} introduces a trial-and-error fine-tuning approach for faster LLM inference. However, the high cost of training domain-specific LLMs for different tasks and device characteristics makes this approach impractical for edge deployment. ToolLLM~\cite{qin2024toolllm} employs a tree-based selection mechanism to reduce the number of tools required for function execution. Unfortunately, in cases when the entire toolset needs to be enumerated, ToolLLM suffers from increased latency and energy consumption. Moreover, Less-is-More~\cite{paramanayakam2024less} proposes an LLM-based tool recommender to identify and prefetch potential tools. However, this method relies on a three-level offline clustering mechanism, creating scalability issues. 
Recent work on efficient function-calling, such as Gorilla \cite{patil2025gorilla} and TinyAgent \cite{erdogan2024tinyagent}, leverages Retrieval-Augmented Generation (RAG) and a transformer-based classifier, respectively, to fetch relevant APIs. However, these techniques might lack generalizability across diverse function spaces, limiting their applicability to dynamic and sustainability-aware function calling at the edge.

Adaptive carbon-aware scheduling techniques have been explored in edge-cloud infrastructures to mitigate the environmental impact of AI workloads. DynamoLLM~\cite{stojkovic2024dynamollm} introduces a predictive scheme based on query complexity to dynamically adjust resource allocation. While it reduces the carbon footprint, it does not fully account for \textit{CI} fluctuations. Moreover, complexity classification remains a non-trivial step, particularly in function-calling, requiring domain-specific refinement. GreenScale~\cite{kim2023greenscale} uses dynamic scheduling across a cloud-edge cluster based on \textit{CI} levels and workload characteristics.
Last, LSCEA-AIoT \cite{song2024carbon} adapts energy-efficient data acquisition and task offloading optimization to achieve reduced carbon emissions.
Unfortunately, these techniques focus more on workload scheduling and task migration over direct optimization of edge resources, particularly neglecting on-device LLM efficiency. CarbonCall differentiates itself by introducing a sustainability-aware function-calling mechanism that integrates dynamic tool selection, real-time power adaptation, and mixed-quality LLMs to minimize operational emissions while ensuring efficient function execution at the edge.

\begin{figure*}
    \centering
    {\includegraphics[width=1.0\linewidth, clip]{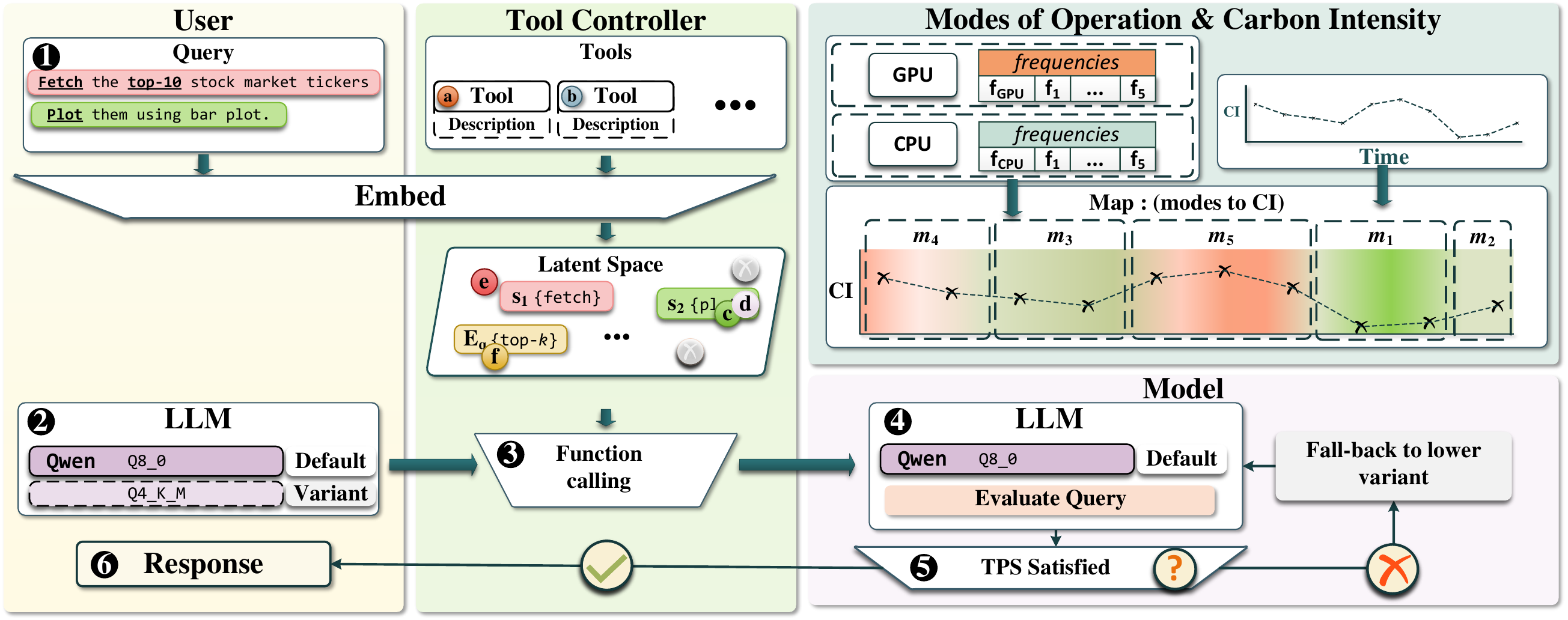}}
    \caption{A high-level overview of CarbonCall.}
    \label{fig:overview}
    \vspace{-5pt}
\end{figure*}

\section{methodology}

CarbonCall optimizes LLM-based function calling on edge devices through carbon-aware execution, as shown in Figure~\ref{fig:overview}. The workflow begins with tool selection, where a shared embedding model converts both the query and tool descriptions into vectors for distance-based retrieval. The LLM then structures function calls, while the system configures the edge server’s operating mode, selecting a power limit ($P_{max}$) based on forecasted carbon intensity. After tool execution, the LLM evaluates the results. If throughput drops below the tokens-per-second threshold, the system switches to a more quantized LLM variant to sustain performance within power constraints.

\subsection{Operational carbon footprint}

The carbon footprint of an edge device is determined by two primary factors: energy consumption and carbon intensity (CI) of the electricity source. Energy consumption, denoted as $E$, is measured in kilowatt-hours (kWh) and represents the total amount of electrical energy used by the device. Carbon intensity, denoted as $CI$, is measured in grams of $CO_2$ per kilowatt-hour ($gCO_2/kWh$) and indicates the amount of carbon dioxide emissions associated with generating one unit of electricity. The total operational carbon footprint of an edge server, represented as $CF$, is given by Equation~\ref{eq:cf}:
\begin{equation}\label{eq:cf}
    CF = E \times CI
\end{equation}
where $CF$ is measured in grams of $CO_2$ ($gCO_2$). Equation~\ref{eq:cf} shows that the total carbon footprint of a system is directly influenced by both its power consumption and the carbon intensity of the electricity grid at the time of operation.

CI varies with the energy mix used by the power grid. Fossil fuels, such as coal (820 $gCO_2/kWh$) and natural gas (490 $gCO_2/kWh$), produce high emissions, while renewable sources like wind (11 $gCO_2/kWh$) and solar (41 $gCO_2/kWh$) have significantly lower impact~\cite{gupta2022act}. As a result, the same energy consumption can lead to different carbon footprints depending on the source. Moreover, CI fluctuates throughout the day due to changes in energy source. Solar power peaks at midday but drops at night, while wind energy depends on weather conditions. This variability means that even if an edge server's energy use remains constant, its carbon footprint changes over time, depending on the grid’s real-time energy mix.

\subsection{Dynamic tool selection}\label{sec:tool_selection}

Dynamic tool selection is essential for enabling efficient function calling in resource-constrained environments. Since edge devices rely on smaller LLMs with fewer parameters (e.g., 1.5B, 3.8B, 7B), these models often lack the robust function-calling capabilities found in larger-scale LLMs~\cite{erdogan2024tinyagent}. Prior studies have shown that when presented with a large number of tools, these models struggle to identify the correct one, leading to higher failure rates~\cite{paramanayakam2024less}. To overcome these challenges, CarbonCall introduces a dynamic tool selection mechanism that efficiently retrieves and ranks only the most relevant tools for a given query. Unlike offline clustering-based methods~\cite{paramanayakam2024less}, which incur high computational costs and scalability concerns, CarbonCall dynamically selects tools using vector embeddings, fast similarity search, and re-ranking.

To determine the most relevant tools for a given user query $q$, CarbonCall first splits the query into sentences $S = \{s_1, s_2, \dots, s_m\}$, ensuring that individual components of complex queries are processed separately. Each sentence $s_i$ and each tool $t_j$ in the toolset $T = \{t_1, t_2, \dots, t_N\}$ are embedded into a shared vector space using a sentence-transformer model optimized for efficient retrieval~\cite{vergou2023readability}:
\begin{equation}
\mathbf{s}_i = \text{Enc}(s_i), \quad \mathbf{t}_j = \text{Enc}(t_j) \quad \forall i \in [m], \quad \forall j \in [N]
\end{equation}
where $ [m] = \{1, \dots, m\} $ and $ [N] = \{1, \dots, N\}$. $Enc$ is the embedding function that converts text into high-dimensional vector representations. These precomputed tool embeddings are stored in FAISS~\cite{douze2024faiss} for efficient nearest-neighbor retrieval. At query time, the query is split into sentences, embedded, and compared against all tools using cosine similarity. Specifically, to efficiently retrieve the most relevant tools, we compute the maximum similarity score for each tool $t_j$ across all sentences $s_i$ of the query:
\begin{equation}
\text{Score}(t_j) = \max_{i \in [m]} \text{Sim}(\mathbf{s}_i, \mathbf{t}_j), \quad \text{Sim}(\mathbf{s}_i, \mathbf{t}_j) := \frac{\mathbf{s}_i \cdot \mathbf{t}_j}{\|\mathbf{s}_i\| \|\mathbf{t}_j\|}
\end{equation}
 
FAISS selects the top-$k$ tools by ranking them based on their scores and returning the $k$ highest-scoring ones. However, FAISS relies solely on precomputed vector embeddings, which may not always capture small differences in meaning between the query and tool descriptions. This can lead to incorrect tool selections~\cite{patil2025gorilla}, especially when multiple tools have similar descriptions or when the query contains ambiguous phrasing that requires deeper contextual understanding. 

To improve accuracy, CarbonCall applies a Cross-Encoder model after FAISS retrieval. Unlike FAISS, which treats the query and tool descriptions independently, the Cross-Encoder jointly processes both in a transformer-based architecture. By comparing each tool in the top-$k$ set
directly with the query in \emph{full context}, the Cross-Encoder captures deeper semantic relationships, enabling more accurate re-ranking based on their \emph{actual} relevance rather than just numerical similarity scores. By integrating this re-ranking step, CarbonCall corrects misalignments in the initial FAISS ranking, ensuring that the selected tools best match the intent of the query. This approach is particularly useful when dealing with ambiguous queries and functionally overlapping tools. Additionally, instead of relying on a fixed similarity threshold, CarbonCall employs an adaptive ranking strategy to determine the number of tools passed to the LLM. If the highest-ranked tool (in the top-$k$ set) is significantly more relevant than the second-best, only that tool is selected to reduce computational overhead. However, if their similarity scores are close, multiple tools are included to handle query complexity and reduce failure rates. 

To further enhance tool selection beyond similarity-based retrieval, CarbonCall also integrates Named Entity Recognition (NER) and keyword-based mappings~\cite{li2020survey}. While the Cross-Encoder refines tool rankings, it may still overlook tools that are relevant based on specific entities or domain-specific terms (e.g., location names) present in the query. To address this, CarbonCall extracts key entities from the user query and maps them to a predefined set of associated tools.

\subsection{Operating modes}\label{sec:operating_modes}

Carbon-aware function calling requires solutions that not only optimize execution but also dynamically adjust power usage based on environmental factors, particularly carbon intensity (CI). CarbonCall’s dynamic tool selection (Section~\ref{sec:tool_selection}) improves efficiency by reducing tool candidates, which lowers power consumption and enhances response accuracy~\cite{paramanayakam2024less}.
However, since carbon footprint is tied to power consumption, real-time CI-based power management is necessary.

To achieve this, CarbonCall employs a lookup table (LUT) of discrete hardware configurations, each corresponding to a specific maximum power consumption level for the edge server. These predefined configurations enable real-time adaptation, reducing operational emissions while maintaining performance efficiency. By selecting the appropriate mode based on both LLM processing demands and CI, CarbonCall can further reduce the environmental impact of function execution (Section~\ref{sec:runtime}). In our implementation on the NVIDIA AGX Orin board, we define five operating modes, where the maximum power limit is gradually constrained from $45W$ down to $28W$ (Table~\ref{tab:operating_modes}). This approach allows for fine-grained control over the device's power consumption, providing flexibility in managing energy use under different workload and CI conditions. To select these values, we conducted an analysis of on-device \emph{tokens-per-second (TPS)} throughput. We found that power caps below $28W$ severely degrade TPS, causing significant response delays in LLM execution. Therefore, while lower power limits could further reduce power consumption, they would introduce unacceptable latencies, limiting the practicality of real-time function execution. By dynamically selecting an appropriate mode from this list based on real-time CI values, CarbonCall effectively caps power consumption without requiring step-by-step frequency adjustments at runtime, which would introduce unnecessary overhead due to the large configuration space.

\begin{table}[H]
\centering
\caption{NVIDIA AGX Orin operating modes.}
\label{tab:operating_modes}
\resizebox{0.8\columnwidth}{!}{
\begin{tabular}{|l|c|c|c|c|}
\hline
\textbf{$m_i$} & $\mathbf{f_{\text{CPU}}}$ & $\mathbf{f_{\text{GPU}}}$ & $\mathbf{f_{\text{mem}}}$ & $\mathbf{P_{\text{max}}}$ \\ \hline
1 & 2.2GH & 1.3GH & 3.1GH & 45W \\
2 & 2.1GH & 1.2GH & 3.1GH & 42W \\
3 & 1.8GH & 1.0GH & 3.1GH & 37W \\
4 & 1.6GH & 918MH & 3.1GH & 33W \\
5 & 1.2GH & 714MH & 3.1GH & 28W \\
\hline
\end{tabular}}
\vspace{-10pt}
\end{table}

\subsection{Quantized LLMs for low-power modes}
In the previous section (Section~\ref{sec:operating_modes}), we introduced discrete operating configurations to cap the maximum power consumption of the edge server. However, reducing CPU and GPU frequencies impact TPS, causing significant response delays. Since function calling in edge environments requires both energy efficiency and responsiveness, maintaining an acceptable TPS under power constraints is important.

To address this issue, CarbonCall integrates quantized LLM variants from the same model family, applying the mixed-quality model paradigm to LLM execution~\cite{berkeley-function-calling-leaderboard}. Instead of using a single LLM across all power modes, CarbonCall dynamically switches to lower-precision, quantized versions when the device operates under low-power constraints. Compared to their full-precision counterparts, the lower-bit precision in quantized LLMs reduces memory and compute requirements while maintaining acceptable accuracy, which in turn enables higher TPS and faster inference at the same CPU/GPU frequencies and power limits. In CarbonCall, we utilize two quantization formats:
\begin{inparaenum}[(1)]
    \item Q8 (8-bit quantization): Retains most of the model’s numerical precision while reducing memory footprint and computational overhead.
    \item Q4\_K\_M (4-bit quantization with K-bit activations and M-bit weights): Further compresses the model, providing a significant boost in efficiency at the cost of a slight reduction in accuracy.
\end{inparaenum}
These optimizations are widely adopted in LLM inference engines, such as HuggingFace, which support the execution of models like Hermes2, Qwen2, or Llama3.1, at various quantization levels.

However, as quantization levels decrease, these models tend to exhibit lower function-calling accuracy~\cite{paramanayakam2024less}. Highly compressed models struggle with more complex tool selection tasks, leading to higher failure rates when identifying the correct API function. This highlights the importance of CarbonCall’s tool selection mechanism (Section~\ref{sec:tool_selection}), which compensates for the loss in LLM accuracy by ensuring that only the most relevant tools are retrieved and passed to the model.
By integrating quantized LLMs with robust tool selection, CarbonCall ensures that function calling remains effective even at lower power configurations.

\subsection{Runtime}\label{sec:runtime}

This section describes how operating modes and quantized LLMs are dynamically managed during runtime. CarbonCall integrates the 24-hour $CI$ forecast for the electricity grid, as presented in~\cite{maji2022carboncast}, to adjust the device’s operational power thresholds in real time. By leveraging carbon-aware power scaling, CarbonCall ensures that function execution remains both energy-efficient and responsive, adapting to variations in environmental impact and system performance.

To determine the optimal power usage strategy, CarbonCall analyzes the predicted $CI$ values over the next 24-hour period, identifying both the minimum and maximum values, denoted as $CI_{min}$ and $CI_{max}$, respectively. At the lowest $CI_{min}$, the environmental impact of power consumption is minimal. Therefore, CarbonCall sets the edge device to its highest power configuration $m_1$ (Table~\ref{tab:operating_modes}), enabling execution at maximum power levels while maintaining efficiency. Conversely, at peak $CI_{max}$, CarbonCall transitions the device to its lowest power mode $m_5$, ensuring that energy consumption is minimized during periods of high environmental impact. For intermediate $CI$ range, CarbonCall gradually adjusts the operating mode by mapping $CI$ values evenly across the available power configurations. This ensures a smooth and adaptive power transition throughout the day, preventing sudden shifts in system performance while aligning execution with carbon-aware energy policies. To avoid frequent, unnecessary power mode changes, CarbonCall updates the maximum power threshold only when the $CI$ changes by at least 10\% of the predicted range. This prevents back-and-forth transitions between power states, ensuring stable operation and minimizing system overhead.

Additionally, CarbonCall continuously monitors the TPS throughput of the LLM execution. Initially, the system runs the Q8 quantized variant to balance performance and efficiency. However, as CPU and GPU frequencies decrease, the TPS can drop significantly, leading to longer response times. To mitigate this, we employ a performance threshold mechanism. If the TPS falls below 80\% of the initial value, CarbonCall transitions to the Q4\_K\_M variant of the same LLM family, ensuring TPS remains within an acceptable range even at lower power modes.

A key challenge in managing runtime model transitions is the overhead associated with loading a new LLM variant. Switching models too frequently can introduce significant delays, leading to oscillations in execution efficiency (a ``pendulum effect''). To prevent this, we apply a 10-minute moving average observation window before making a transition decision. By averaging TPS over a 10-minute period, CarbonCall avoids rapid fluctuations in model switching, ensuring stable execution while adapting to power and carbon constraints.

In summary, CarbonCall integrates real-time CI forecasts, adaptive power management, and quantized LLM transitions to optimize function execution for both sustainability and performance. By dynamically adjusting power thresholds and model variants, it effectively reduces the carbon footprint while maintaining a high-quality user experience in edge-based AI deployments, as shown in Section~\ref{sec:evaluation}.

\section{Experimental results}\label{sec:evaluation}

We evaluated CarbonCall using the BFCL~\cite{berkeley-function-calling-leaderboard} and GeoEngine~\cite{singh2024llm} benchmarks. Queries in BFCL require single function calls per sub-question, while GeoEngine involves multiple sequential function calls, making them challenging. To ensure a diverse difficulty distribution and realistic user-based input, we combined queries from both benchmarks. All experiments were conducted on an NVIDIA Jetson AGX Orin board.
which offers \begin{inparaenum}[(\bgroup\bfseries i\egroup)]
    \item 2048-core NVIDIA Ampere architecture GPU with 64 Tensor Cores;
    \item 12-core ARM Cortex-A78 CPU with 3 clusters of 4 cores each; and
    \item 64GB LPDDR5 memory.
    With these specs, the Jetson AGX Orin can perform up to 270 TOPS, capped at 60W, making it a suitable platform to evaluate small-sized LLMs in edge settings.
\end{inparaenum}

For our evaluation, we selected the Q8 and Q4\_K\_M quantized variants of three LLM families:
\begin{inparaenum}[(i)]
\item Hermes2-Pro-8B,
\item Llama3.1-8B, and
\item Qwen2-7B.
\end{inparaenum}
Each model was deployed on the edge server for five consecutive days to assess the impact of CI fluctuations on function calling. We simulated realistic workloads by generating query combinations from a random mix of BFCL and GeoEngine benchmarks, ensuring diverse function-calling scenarios. This allowed us to analyze LLM performance under varying energy constraints while also expanding the pool of available tools for execution.

We compared CarbonCall against: \begin{inparaenum}[(\bgroup\bfseries i\egroup)] 
    \item \textbf{Default} method, which invokes the LLMs with all the available functions and no other optimization; 
    \item \textbf{Gorilla}~\cite{patil2025gorilla}, which filters functions by directly matching the queries and the function descriptions;
    \item Less-is-More \textbf{(LiS)}~\cite{paramanayakam2024less}, which uses the LLM's reasoning capabilities to generate its own set of functions per query;
    \item \textbf{LiS*}, a modified version of \textbf{LiS}, which changes the operation modes according to the carbon intensity while running the same quantized variant.
\end{inparaenum}

We evaluated our method by running experiments on all these approaches and comparing the normalized values of:
\begin{inparaenum}[(\bgroup\bfseries i\egroup)]
    \item average latency (\textbf{T$_{Norm}$});
    \item average power consumption (\textbf{P$_{Norm}$});
    \item average token per second (\textbf{TPS$_{Norm}$}); and
    \item average carbon emission per query (\textbf{CF$_{Norm}$}).
\end{inparaenum}

\begin{figure}[]
    \centering
    \resizebox{1\columnwidth}{!}{\includegraphics[width=\linewidth, clip]{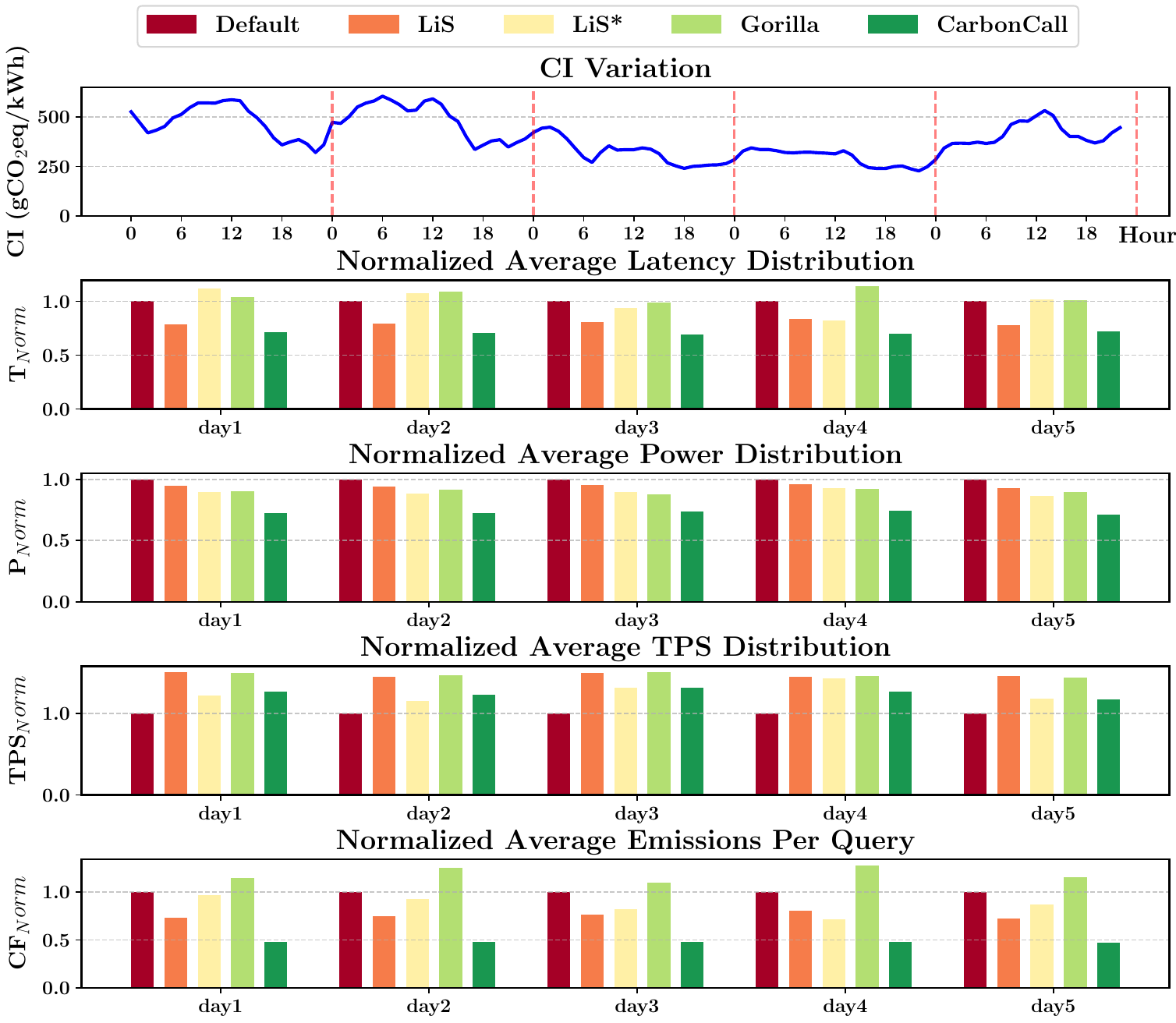}}
    \caption{Week 1: Performance evaluation of Hermes2-pro-8B.}
    \label{fig:res-hermes}
\end{figure}

\begin{figure}[]
    \centering
    \resizebox{1\columnwidth}{!}{\includegraphics[width=\linewidth, clip]{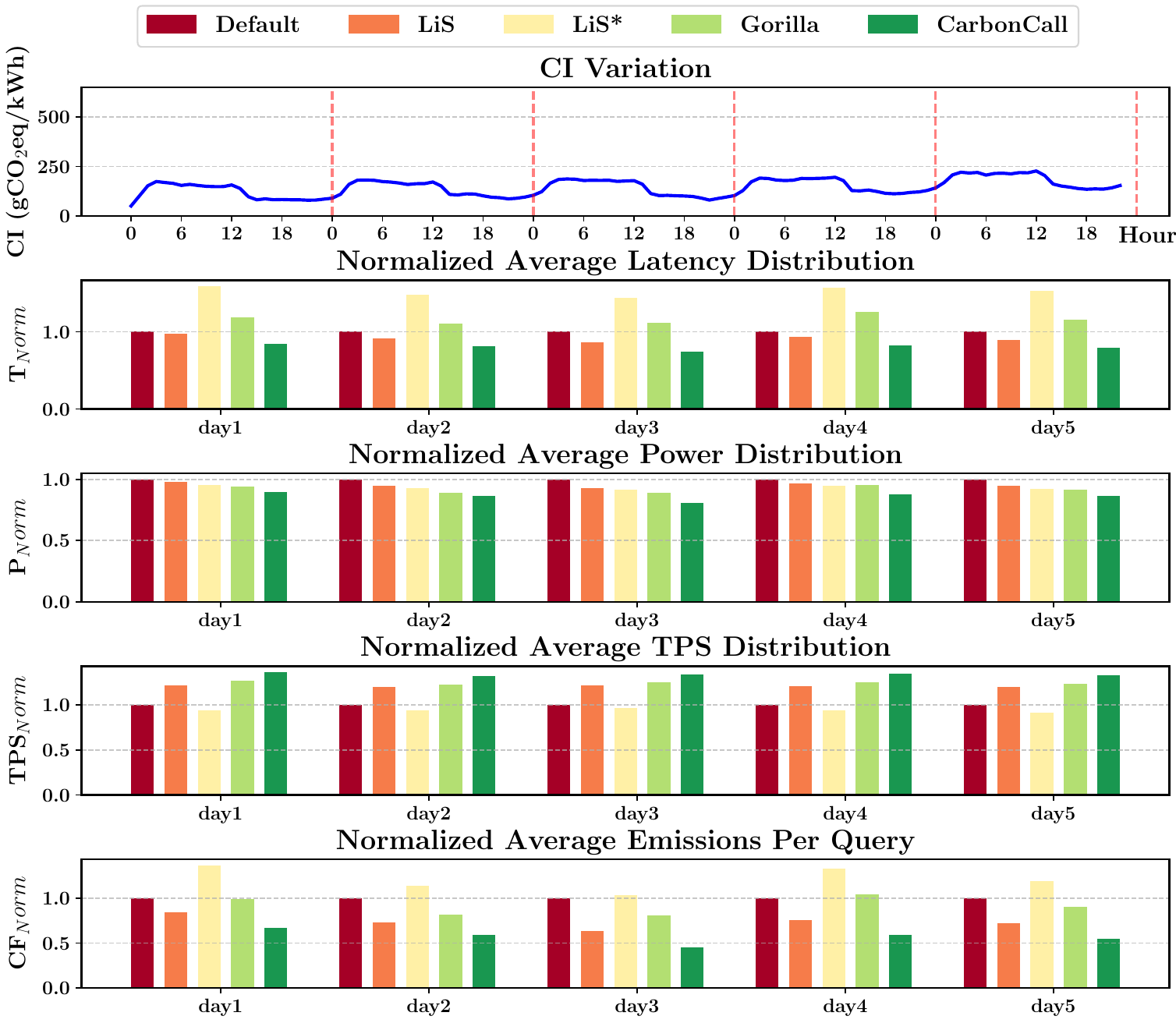}}
    \caption{Week 2: Performance evaluation of Llama3.1-8B.}
    \label{fig:res-llama}
\end{figure}

\begin{figure}[]
    \centering
    \resizebox{1\columnwidth}{!}{\includegraphics[width=\linewidth, clip]{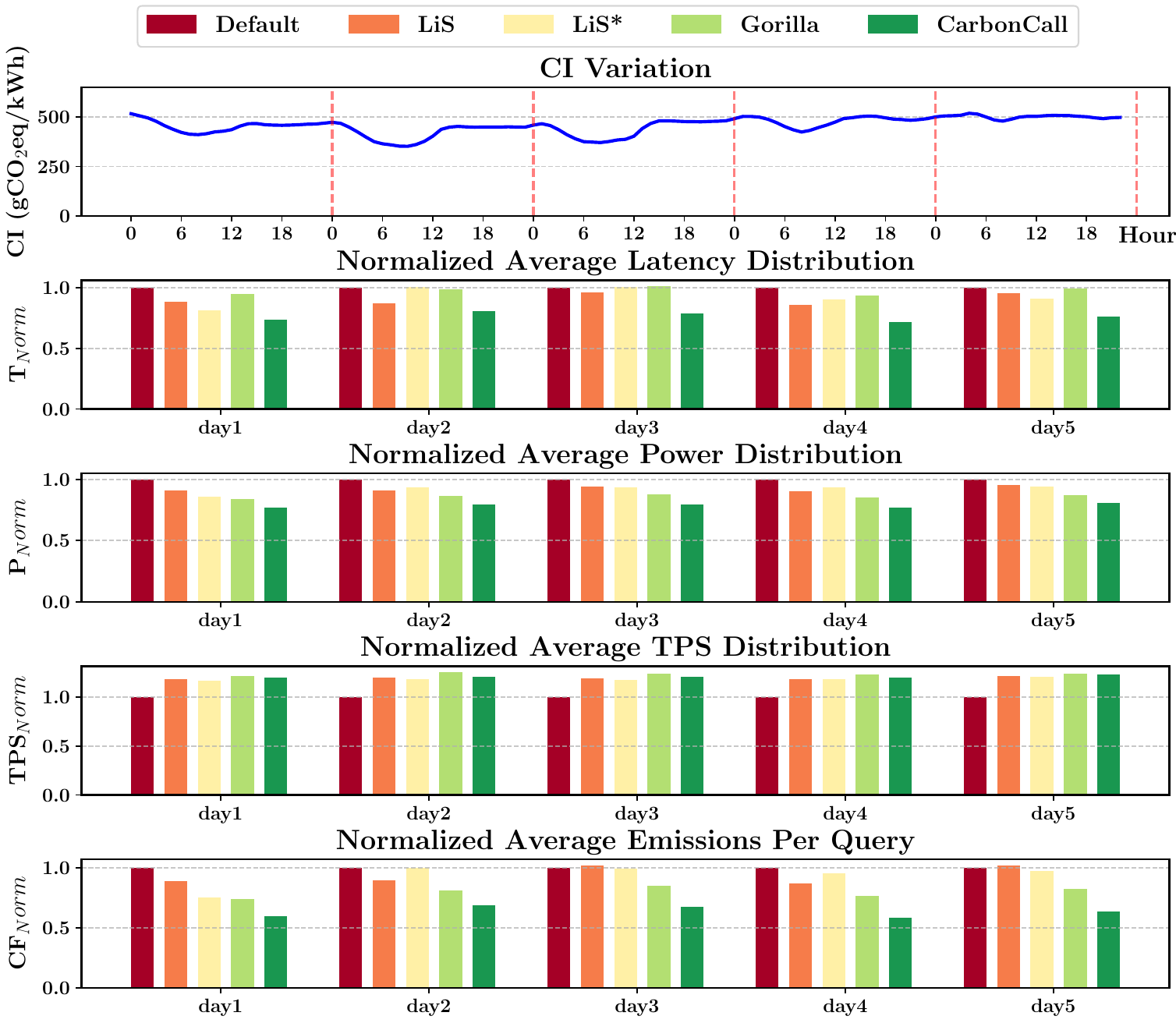}}
    \caption{Week 3: Performance evaluation of Qwen2-7B.}
    \label{fig:res-qwen}
\end{figure}

\begin{figure}
    \centering
    \resizebox{1\columnwidth}{!}{\includegraphics[width=\linewidth, clip]{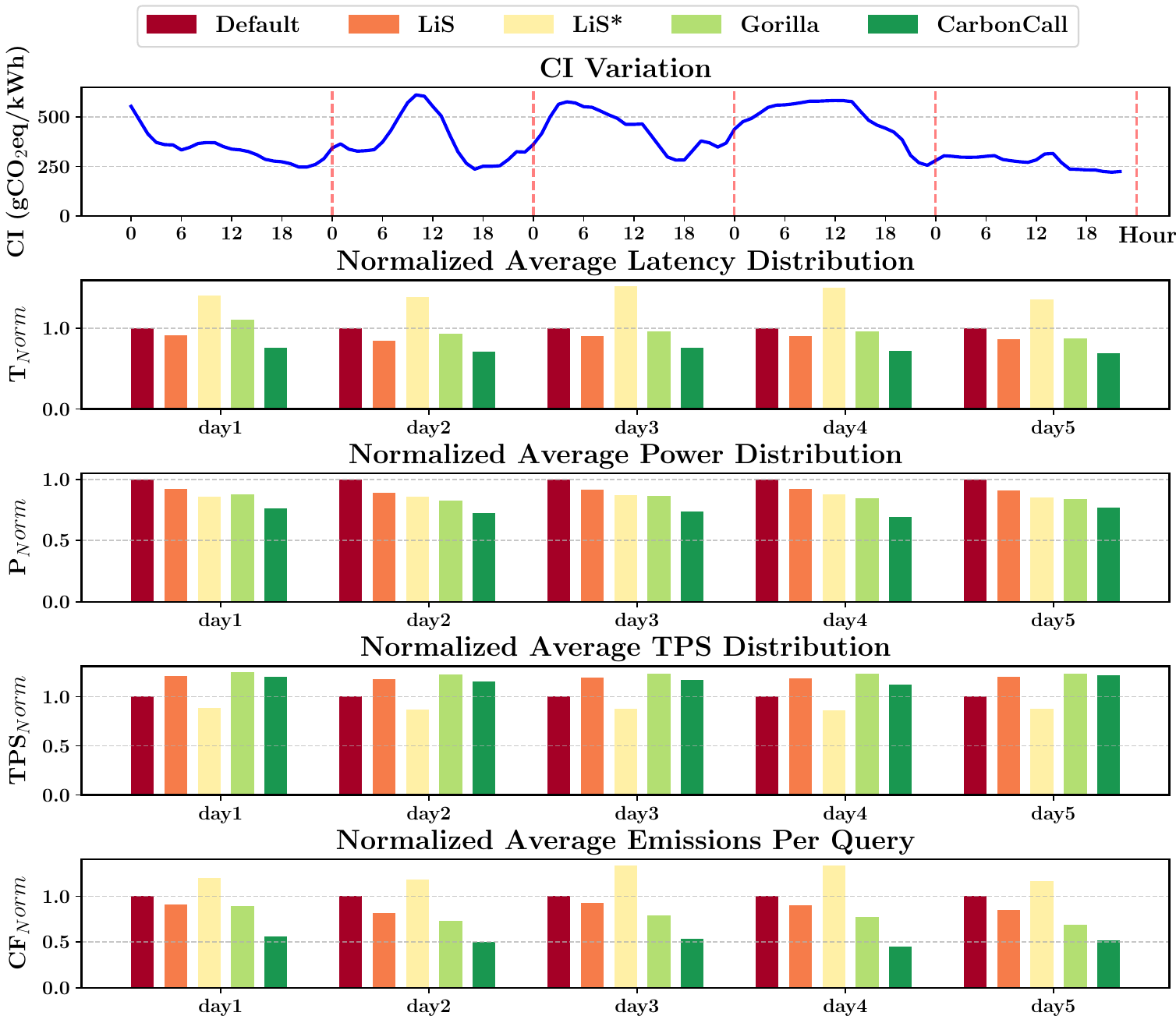}}
    \caption{Week 4: Performance evaluation os Qwen2-7B.}
    \label{fig:res-qwen2}
\end{figure}

Figures \ref{fig:res-hermes}-\ref{fig:res-qwen2} present the results across four weeks with varying carbon intensity (CI) levels. Figure \ref{fig:res-hermes} illustrates the performance of Hermes2-Pro-8B in week 1, where $CI$ fluctuates between 220 gCO$_2$/kWh and 610 gCO$_2$/kWh with moderate to high variability. On average, CarbonCall reduced power consumption (P$_{Norm}$) by 28\%, execution time (T$_{Norm}$) by 30\%, and carbon emissions (CF$_{Norm}$) by 52\%, while improving TPS (TPS$_{Norm}$) by 25\% compared to the default method. Despite operating under power constraints, our method still achieved 9\% lower T$_{Norm}$ than LiS. LiS and Gorilla showed similar TPS$_{Norm}$, with 47\% higher TPS$_{Norm}$ than the default method, while CarbonCall, operating at lower power levels, reached 25\% higher TPS$_{Norm}$. Among all methods, Gorilla had the highest emissions, whereas LiS and LiS* fell between the default method and CarbonCall.

The Llama3.1-8B was evaluated in week 2, where the \textit{CI} varied between 70 gCO$_2$/kWh to 230 gCO$_2$/kWh with moderate fluctuations. The results are shown in Figure~\ref{fig:res-llama}. Our method achieved 20\% and 12\% improved T$_{Norm}$, 14\% and 10\% less P$_{Norm}$, and 47\% and 13\% less emissions compared to default and LiS, respectively. LiS* and Gorilla incurred higher delays, with LiS* being the worst. In this experiment, our method resulted in the highest TPS. The higher TPS is because, as the operated \textit{CI} range is smaller and the CI mostly falls in the higher end of the range, CarbonCall spent less time in the lower power modes, resulting in higher average TPS.

For the Qwen2-7B model, Figure~\ref{fig:res-qwen} depicts a week (week 3) with low \textit{CI} fluctuations between 350 gCO$_2$/kWh and 520 gCO$_2$/kWh, while Figure~\ref{fig:res-qwen2} shows week 4 with high fluctuations from 200 gCO$_2$/kWh to 620 gCO$_2$/kWh. For both week 3 and week 4, our method achieved 16\% and 20\% better T$_{Norm}$, 14\% and 19\% less P$_{Norm}$, 37\% and 47\% less emissions, respectively, compared to LiS. Due to lower \textit{CI} variations in week 3, LiS* performed similarly to LiS. In contrast, during week 4, higher \textit{CI} variations led to LiS* yielding the worst metrics. In both weeks, our method achieved TPS comparable to LiS, while Gorilla achieved the highest TPS.

Overall, our method reduced carbon emissions by up to 52\% and 47\%  compared to the Default and LiS baselines, respectively, while lowering power consumption by 30\% and 14\%, with a manageable TPS deficit in some cases. Despite the reduced power, execution time improved in all experiments, demonstrating CarbonCall’s ability to cut emissions without sacrificing performance. Although Gorilla has computational complexity comparable to our approach, it resulted in higher emissions, at times the highest among all methods. This is because, while Gorilla performs well on BFCL \cite{berkeley-function-calling-leaderboard}, it struggles with GeoEngine \cite{singh2024llm} queries, leading to longer execution times and increased emissions.

\begin{figure}[]
    \centering
    \resizebox{1\columnwidth}{!}{\includegraphics[width=\linewidth, clip]{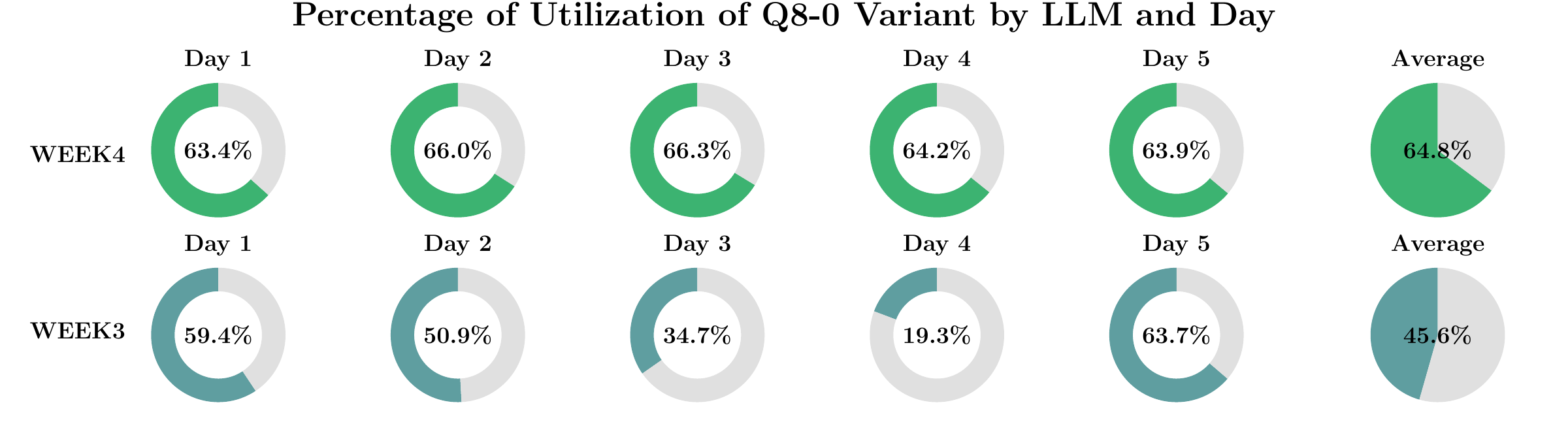}}
    \caption{Utilization of the Q8 variant across two Qwen2-7B week experiments conducted under different $CI$ variations.}
    \label{fig:res-qwen-util}
\end{figure}

Based on the Qwen2-7B experiments for weeks 3 and 4, we compared the \textbf{model variant utilization} in Figure \ref{fig:res-qwen-util}, which shows the utilization of the Q8 variant for each weekday. In week 3, which has lower CI fluctuations, Q8 was used for 64.8\% of the queries on average, while in week 4 it was used 45.6\% of the time. Overall, we observe that days with higher CI generally led to lower utilization of the Q8 variant. However, we note that the lowest CI days did not necessarily correspond to higher Q8 utilization. This is because variant selection depends on TPS and, consequently, on query complexity. Despite this small variation, overall, in the lower fluctuating CI ranges, the Q8 variant was used more.

\section{Conclusion}
We presented CarbonCall, a sustainability-aware framework for LLM execution on edge devices. By combining dynamic tool selection, carbon-aware execution, and quantized LLM adaptation, CarbonCall minimizes power consumption and emissions while preserving response speed. Evaluation shows 52\% lower emissions, 30\% lower power use, and better execution time. Compared to existing methods, CarbonCall achieves higher energy efficiency, making it a practical solution for sustainable agentic AI at the edge. 

\section{Acknowledgement}
This work is supported by grant NSF CCF 2324854. Any opinions, findings, and conclusions or recommendations expressed in this material are those of the authors and do not necessarily reflect the views of the National Science Foundation.

\bibliographystyle{IEEEtran}
\bibliography{ref}
\end{document}